# Advanced LSTM Neural Networks for Predicting Directional Changes in Sector-Specific ETFs Using Machine Learning Techniques

Rifa Gowani
*New York University*
New York, New York
United States
RifaGowani@nyu.edu

Zaryab Kanjiani
*University of North Texas*
Denton, Texas
United States
ZaryabKanjiani@my.unt.edu

*Abstract*— **Trading and investing in stocks for some is their full-time career, while for others, it's simply just a supplementary income stream. Universal among all investors is the desire to turn a profit. The key to achieving this goal is diversification. Spreading your investments across sectors is the key to profitability and maximizing returns. This study aims to gauge the viability of machine learning methods to practice the principle of diversification to maximize your portfolio returns. To test this, this study tests the Long-Short Term Memory (LSTM) model across 9 different sectors and upwards of 2,200 stocks using Vanguard's sector-based ETFs. Across all sectors, the R-squared value showed very promising results, with an average of .8651 and a high of .942 with the VNQ ETF. These findings suggest that the LSTM model is a capable and viable model for accurately predicting directional changes among various industry sectors and can help investors diversify and grow their portfolios.**

## I. INTRODUCTION

An important principle in investment banking is diversification. By spreading investments across various sectors and industries, investors mitigate risk and increase their potential for long-term stable returns. This strategy allows investors to cut their losses and ensure that a downturn in one sector does not impact their entire portfolio [1]. Within this expansive market, sectoral analysis becomes key to understanding specific segments' nuanced movements and trends. Sectoral investing allows the public to invest in different parts of the economy, diversifying their investment strategy. This approach lets investors reap the benefits of each sector while remaining protected against sector-specific downtrends.

More and more investors are increasingly turning to ETFs to practice these diversification principles. Vanguard's ETFs are popular in this regard as they offer access to a selection of stocks across sectors, from technology, finance, healthcare, communication, and many more. These ETFs allow investors to quickly spread their investments to many sectors rather than going through the headache of individually examining stocks. This quality and convenience make them well-suited for in-depth analytical assessments.

In the last few decades, researchers have shown keen interest in stock market prediction using machine learning techniques [2]. The main reason researchers have opted for the use of machine learning techniques to predict stock prices is because they are efficient, effective, and accurate in predicting the market value of a stock [3]. These predictions tend to be close to the real tangible value and have proven to be a structured, systematic approach to predicting the price of a stock [3].

Insights gained from researching sectoral stock prediction using machine learning models can significantly aid investors in diversifying their portfolios. In this study, a Long Short-Term Memory (LSTM) network, a machine learning model, was employed to forecast stock prices. In contrast to traditional models such as Autoregressive Integrated Moving Average (ARIMA) and Generalized Autoregressive Conditional Heteroskedasticity (GARCH), LSTMs do not necessitate stationary data. They are capable of handling complex, nonlinear relationships [4]. Unlike classical machine learning models, LSTMs eliminate the necessity for extensive feature engineering to address temporal dependencies. By effectively addressing the vanishing gradient problem, LSTMs outperform traditional Recurrent Neural Networks (RNNs), making them suitable for long-term forecasting like what is tested in this project [5].

## II. METHODOLOGY

The dataset utilized for analysis was historical stock data gathered from the Yahoo Finance API. This data spanned from January 1, 2012, to December 21, 2022, and included information for nine ETFs, consisting of upwards of 2200 stocks. The ETFs analyzed were the Vanguard Sector-based ETFs: The Vanguard Information Technology ETF (VGT), Vanguard Financials ETF (VFH), Vanguard Consumer Discretionary ETF (VCR), Vanguard Health Care ETF (VHT), Vanguard Communication Services ETF (VOX), Vanguard Industrials ETF (VIS), Vanguard Energy ETF (VDE), Vanguard Real Estate ETF (VNQ), and Vanguard Utilities ETF (VPU). All the stocks listed on these funds were from the NYSE or NASDAQ and were constantly updated. The Vanguard Sector-based ETFs were selected for their reliable performance, comprehensive sector coverage, and the advantage of having a single, uniform provider, ensuring no stock overlap.

Due to the various blank values in the dataset, all data was first pre-processed. Initially, these blank values were identified and removed using the dropna method from the Pandas library, which effectively scans for and eliminates any NaN (Not a Number) entries. This approach was chosen over imputing missing data to maintain the sequential integrity of time series data, such as stock prices, where each record is chronologically ordered. This step is particularly important for LSTM networks, a specialized form of Recurrent Neural Networks (RNNs) designed to handle data sequences and capture long-term dependencies. LSTMs exhibit high sensitivity to the quality of input data, underscoring the necessity of employing a comprehensive and precise dataset. Following data cleansing, graphical representations of the 100-day and 200-day moving averages were generated to facilitate a more comprehensive visualization and comprehension of the underlying trends. These were calculated by:

$$\text{SMA}_n = \frac{P_t + P_{t-1} + P_{t-2} + \cdots + P_{t-(n-1)}}{n} \quad (1)$$

After visualization, the data was split into an 80-20 ratio, with 80% reserved for training the model and 20% for testing its predictions. The iloc method, a method that allows users to select specific rows or columns from a data set, separated the dataset into two DataFrames: data_train for training and data_test for testing. Splitting the data this way gave the model a large and diverse set of historical points and a sufficient timeline to demonstrate its predictive abilities.

### A. Model Architecture

The model was constructed using the Sequential API from Keras, a widely used deep learning framework that simplifies the process of building and training neural networks [6]. The first layer of the model begins with an LTSM layer, which is a Recurrent Neural Network (RNN) capable of learning long-term dependencies. LSTMs are well suited for stock data as they work well over long sequences and time series data. The LTSM layer contains 50 units, which are essential for memory processing and for the model to capture temporal dependencies from the data. This layer is configured with "return_sequences=True" to ensure that the output of this layer will be a sequence of the same length as the input; this is an essential step when stacking multiple LSTM Layers.

The main components of the LSTM network are the cell state ($C_t$), hidden state ($h_t$), input gate ($i_t$), forget gate ($f_t$) output gate ($o_t$), and candidate cell state ($\tilde{C}_t$). These are computed as follows:

Forget Gate:
$$f_t = \sigma(W_f \cdot [h_{t-1}, x_t] + b_f) \quad (2)$$

Input Gate:
$$i_t = \sigma(W_i \cdot [h_{t-1}, x_t] + b_i) \quad (3)$$

$$\tilde{C}_t = \tanh(W_C \cdot [h_{t-1}, x_t] + b_C) \quad (4)$$

Cell State:
$$C_t = f_t * C_{t-1} + i_t * \tilde{C}_t \quad (5)$$

Output Gate:
$$o_t = \sigma(W_o \cdot [h_{t-1}, x_t] + b_o) \quad (6)$$

Hidden State:
$$h_t = o_t * \tanh(C_t) \quad (7)$$

Following the first layer is a dropout layer with a rate of 20%. This dropout layer is an essential regulatory technique used to prevent overfitting. In this common problem, the model learns the training data too well but fails to generalize to new, unseen data [7]. This dropout layer randomly sets 20% of the input units to zero during each update at training time, helping prevent the co-adaptation of neurons and promoting the independence of features.

Following the dropout layer, a second LSTM layer with 60 units is added. This layer is configured with "return_sequences=True" so the sequence continues to the next layer. It's then followed by another dropout layer set at 30% to enhance regularization. This structure is repeated twice: first, with an LSTM layer of 80 units and a 40% dropout rate, and then with an LSTM layer of 120 units and a 50% dropout rate.

The final layer is a dense layer with one unit that outputs one singular value: the predicted stock price. A dense layer containing only one unit was done purposefully as it is well suited for regression tasks where the goal is to predict a continuous value.

### B. Model Compilation

Once the model architecture was wholly defined, it was all compiled. During compilation, two main components were specified: the optimizer and the loss function. The optimizer used in our model was the Adam (Adaptive Moment Estimation) Optimizer due to its computational efficiency and ability to handle sparse gradients during "noisy" problems [8]. The loss function used in our model was the Mean Squared Error (MSE), which is the standard choice for most regression problems as it measures the average of the squares of the errors. This is the difference between the estimated values and the actual value. Mathematically, it is represented as:

$$MSE = \frac{1}{n} \sum_{i=1}^{n} (y_i - \hat{y}_i)^2 \quad (8)$$

### C. Model Training

For training the model, the fit method was used to iterate over the training data for the specified number of epochs. In this study's case, the model was trained for 50 epochs with a batch size of 32. An epoch refers to one complete pass through the entire training dataset and the batch size refers to the number of training examples utilized in one iteration. During this process, the model's performance and loss are closely monitored to adjust internal weights and biases, thereby improving its ability to predict stock prices accurately. By the end of the whole training process, the model has enough knowledge of the temporal dependencies and patterns in the stock price data making it capable of making future price predictions.

### III. RESULTS & DATA

An analysis was performed to analyze how well the model could predict stock prices using LSTM networks across various sectors represented by Vanguard Sector-based ETFs. This analysis includes performance metrics for each sector to determine the accuracy and reliability of the LSTM model in predicting stock prices.

## A. Model Root Mean Squared Error (RMSE)

The Root Mean Squared Error (RMSE) quantifies the average magnitude of the discrepancy between predicted and actual values. Lower RMSE values denote superior model performance. This metric is particularly valuab greater significance to larger errors, rendering it sensitive to outliers.

TABLE I. ROOT MEAN SQUARED ERROR (RMSE) BY SECTOR

| Sector | ETF Sector | Root Mean Squared Error (RMSE) |
|---|---|---|
| Information Technology | VGT | 11.2226 |
| Financials | VFH | 2.0755 |
| Consumer Discretionary | VCR | 8.1826 |
| Health Care | VHT | 5.4498 |
| Communication Services | VOX | 5.6090 |
| Industrials | VIS | 6.8448 |
| Energy | VDE | 6.0787 |
| Real Estate | VNQ | 1.9013 |
| Utilities | VPU | 2.7150 |

Based on the RMSE values, it can be seen that the Real Estate and Financials sectors exhibit lower prediction errors, implying a higher degree of prediction accuracy within these sectors., the Information Technology sector shows the highest RMSE, reflecting greater challenges in achieving accurate predictions.

## B. Mean Absolute Error (MAE) by Sector

The Mean Absolute Error (MAE) quantifies the average magnitude of the absolute variances between the forecasted and actual values. It presents a straightforward measure of prediction precision by demonstrating the average error magnitude without being as susceptible to outliers as RMSE.

| Sector | ETF Symbol | Mean Absolute Error (MAE) |
|---|---|---|
| Information Technology | VGT | 11.2226 |
| Financials | VFH | 2.0755 |
| Consumer Discretionary | VCR | 8.1826 |
| Health Care | VHT | 5.4498 |
| Communication Services | VOX | 5.6090 |
| Industrials | VIS | 6.8448 |
| Energy | VDE | 6.0787 |
| Real Estate | VNQ | 1.9013 |
| Utilities | VPU | 2.7150 |

TABLE II. MEAN ABSOLUTE ERROR (MAE) BY SECTOR

The MAE results are consistent with the RMSE findings, highlighting the Real Estate and Financials sectors as having the lowest errors. High MAE in the Information Technology sector suggests significant prediction inaccuracies.

## C. Additional Metrics by Sector

These additional metrics provide further insights into the model's performance:

- R-Squared ($R^2$): Indicates how well the model explains the variability of the dependent variable. Higher values signify better model fit.
- Mean Absolute Percentage Error (MAPE): Measures prediction accuracy as a percentage error.
- Explained Variance Score: Shows how much of the data variability the model captures. Higher scores indicate better performance.

TABLE III. ADDITIONAL METRICS BY SECTOR

| Sector | ETF Symbol | R-Squared ($R^2$) | Mean Absolute Percentage Error (MAPE) | Explained Variance Score |
|---|---|---|---|---|
| Information Technology | VGT | 0.8895 | 0.1094 | 0.8919 |
| Financials | VFH | 0.9218 | 0.0821 | 0.9302 |
| Consumer Discretionary | VCR | 0.9132 | 0.0979 | 0.9175 |
| Health Care | VHT | 0.7901 | 0.1035 | 0.8638 |
| Communication Services | VOX | 0.9017 | 156859941251053.8 | 0.9317 |
| Industrials | VIS | 0.6810 | 0.1092 | 0.8537 |
| Energy | VDE | 0.9095 | 119367424224013.73 | 0.9369 |
| Real Estate | VNQ | 0.9423 | 0.1303 | 0.9428 |
| Utilities | VPU | 0.8373 | 0.1154 | 0.8378 |

The Real Estate and Financials sectors again show high R-squared values, indicating strong model performance in these sectors. The Energy and Communication Services sectors have unusually high MAPE values, suggesting potential issues with model performance or data quality in these sectors.

## D. Stock Price Prediction

Below are the stock price prediction graphs for each sector. These graphs compare the original stock prices (green line) with the predicted stock prices (red line).

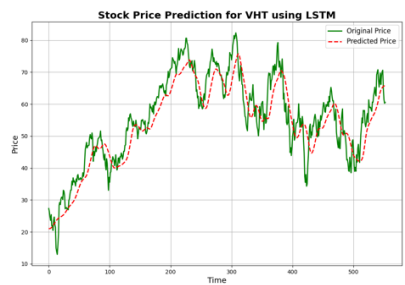

Fig. 1. *Example Stock Price Predictions vs. Actual Prices for VFH ETF Using LSTM Models*

In Figure 1, the graph displays the performance of the VGT ETF. The model captures the upward trend, although there are some discrepancies during periods of rapid price changes. The overall prediction accuracy is good, with the model following actual price movements closely.

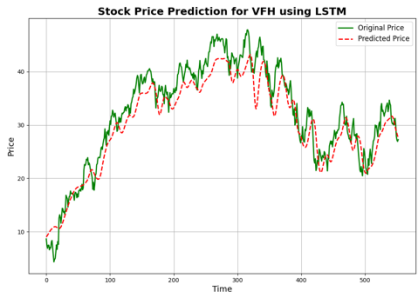

Fig. 2. *Example Stock Price Predictions vs. Actual Prices for VFH ETF Using LSTM Models*

The VFH, represented in Figure 2, shows high accuracy in predictions, as reflected by the close alignment of the predicted and actual price lines, suggesting reliable performance in the financial sector.

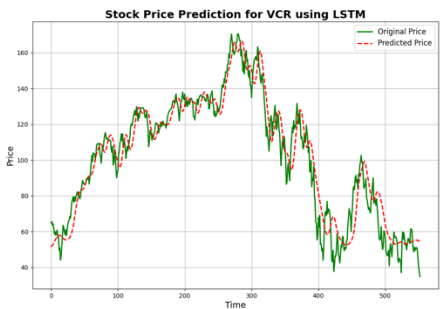

Fig. 3. *Example Stock Price Predictions vs. Actual Prices for VCR ETF Using LSTM Models*

The graph in Figure 3 displays the performance of the VCR ETF. The LSTM model effectively captures the general trends of stock prices, with noticeable deviations during periods of high volatility.

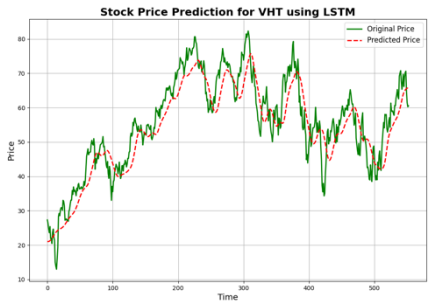

Fig. 4. *Example Stock Price Predictions vs. Actual Prices for VHT ETF Using LSTM Models*

For the VHT, shown in Figure 4, the LSTM model performs moderately well, capturing the general trend but occasionally missing short-term fluctuations, indicating room for improvement in handling rapid price changes.

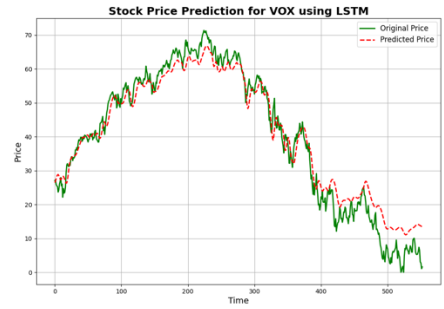

Fig. 5. *Example Stock Price Predictions vs. Actual Prices for VFH ETF Using LSTM Models*

The VOX ETF in Figure 5, shows mixed performance. While the model accurately predicts certain periods, it underperforms during others, especially towards the end of the timeline.

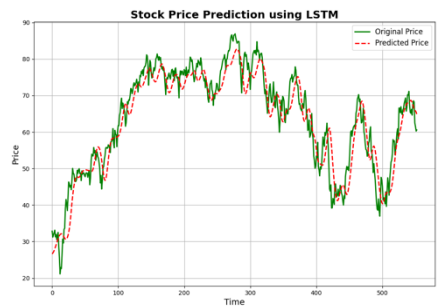

Fig. 6. *Example Stock Price Predictions vs. Actual Prices for VFH ETF Using LSTM Models*

The VIS ETF represented in Figure 6, demonstrates a good fit, with predicted prices reasonably following actual prices. However, some discrepancies, particularly in highly volatile periods, suggest the model could benefit from further tuning.

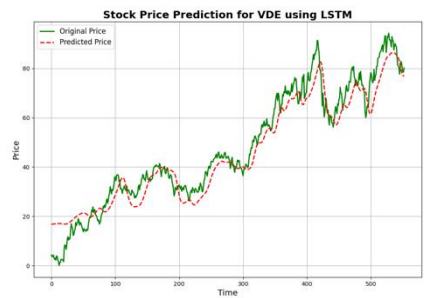

Fig. 7. *Example Stock Price Predictions vs. Actual Prices for VFH ETF Using LSTM Models*

Moving to Figure 7, the Vanguard Energy ETF (VDE) demonstrates the model's efficacy in predicting the overall upward trend and fluctuations, aligning closely with the actual prices. This suggests the model's reliability in this sector.

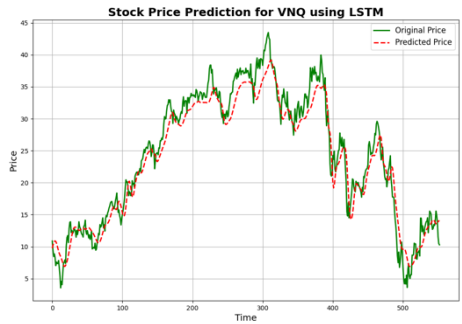

Fig. 8.  *Example Stock Price Predictions vs. Actual Prices for VFH ETF Using LSTM Models*

For the VNQ ETF shown in Figure 8, the model exhibits excellent performance, with predicted prices closely tracking actual prices, indicating robustness in predicting real estate sector prices.

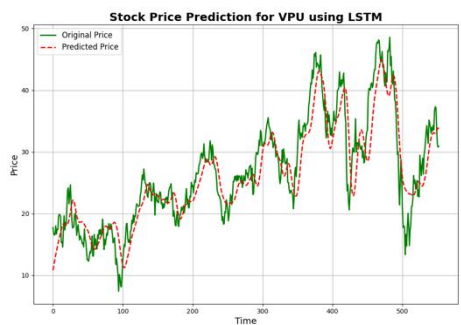

Fig. 9.  *Example Stock Price Predictions vs. Actual Prices for VFH ETF Using LSTM Models*

The VPU ETF in Figure 9, also shows good prediction accuracy, with the predicted prices aligning closely with actual prices, demonstrating the model's effectiveness in capturing trends and fluctuations in the utilities sector.

## IV. Discussion

For this research study, the choice of utilizing an LSTM model was a deliberate and effective decision. Some methods that current researchers have been using to predict stock prices have been the use of traditional statistical methods or classical machine learning methods. Traditional statistical techniques such as Autoregressive Integrated Moving Average or Generalized Autoregressive Conditional Heteroskedasticity are both widely used for time series data as they are effective for capturing linear relationships, but they have regularly been shown to struggle with nonlinear stock patterns and capturing long-term dependencies. The other choice of classical machine learning methods, such as Support Vector Machines or Random forests, are also both heavily utilized in the field of stock price prediction but lacking in the areas of handling time series data or capturing dependencies and temporal patterns [9].

The LSTM model excels in capturing long-term dependencies as its memory cell structure allows it to retain information over long periods. LSTMs are also able to handle nonlinear relationships very effectively without extensive feature engineering, allowing them to adapt to various patterns and trends in stock price. Lastly, the LSTM model is structured in a way suited for sequential data, making it the ideal choice for time series data and future forecasting.

## V. Conclusion

This paper was aimed at testing the viability of machine learning and the employment of investment banking principles to help investors increase portfolio returns. The model used for the purposes of this research study was an intricate and complex LSTM model. The advantages of using this LSTM model were clearly shown in the results it was able to obtain. Looking at the performance of the model, when we focus on the R-Squared value for all sectors, the LSTM model's performance stands out, particularly in the Real Estate Sector ETF where it achieved a high of 0.942, contributing to an impressive average of .8651 across all sectors. The results from the model's directional prediction capabilities are very promising and prove that the LSTM model is effective at accurately predicting the movement of industry sectors. Proving also that it can be used to capitalize on diversification to increase your portfolio returns.